\documentclass[aps,prl,twocolumn,superscriptaddress,groupedaddress, showkeys]{revtex4}  % for review and submission
\usepackage{graphicx}  % needed for figures
\usepackage{dcolumn}   % needed for some tables
\usepackage{bm}        % for math
\usepackage{amssymb}   % for math
\usepackage{graphicx, epsfig, subfigure}

\usepackage{color}

\begin{document}

\title{Quadratic Gravitational Lagrangian with Torsion Can Give Possible Explanations of the Form of Galactic Rotation Curves, of the Amount of Intergalactic Lensings, and of the Accelerating Expansion of the Universe} 

%\author{
%        Y. Yang$^{a}$,  W.B. Yeung$^{b}$\thanks{test}\\
%        $^{a}$Indiana University, Bloomington, IN, USA \\ 
%        $^{b}$Institute of Physics, Academia Sinica, Taipei, Taiwan, ROC
%       }

\author{Y. Yang}
%\email{yi.yang@cern.ch}
\affiliation{Indiana University, Bloomington, IN, USA}
\author{W.B. Yeung}
\thanks{Corresponding author, Phone: 886-22789-8966, Fax: 886-2788-9828}
\email{phwyeung@phys.sinica.edu.tw}
\affiliation{Institute of Physics, Academia Sinica, Taipei, Taiwan, ROC}
\date{\today}

%\footnote{Phone: 886-227898966}\footnote{Fax: 886-27889828}
%\cortext[cor1]{Corresponding author}
%\vspace*{1cm}

\begin{abstract}
The Quadratic Gravitational Lagrangian with torsion provides us with a richer number of solutions than the Einstein-Hilbert Lagrangian does. With proper interpretation, these solutions, together, seem to give good explanations of the form of the galactic rotation curves, of the amount of intergalactic gravitational lensings, and of the accelerating expansion of the Universe.
%The existence of Particle Families can also arise from the existence of these various microscopic metrics endowed to the respective particles 
\end{abstract}


\maketitle

\section{Introduction}
Hermann Weyl once remarked that the most natural gravitational Lagrangian should be quadratic in the Riemann curvature tensor~\cite{Weyl:1919}. 
This remark looks particularly intriguing if we consider the great success of the Yang-Mills way of describing various kinds of interactions which goes with a Lagrangian quadratic in field strength tensors.

The equation of motion, following from the quadratic gravitational Lagrangian was studied, in one way or the other, by many people including Lanczos~\cite{Lanczos:1949}, Stephenson~\cite{Stephenson:1958}, Kilmister~\cite{Kilmister:1962}, Yang~\cite{Yang:1974}, Thompson, Pirani, and Pavelle~\cite{Thomspson:1975}, Hehl and his collaborators~\cite{Hehl:1979}, and Hsu and Yeung~\cite{Yeung:1987}. Their studies didn't attract much attentions in those days because the quadratic Lagrangian didn't reveal much new physics by then. People still prefer the Einstein-Hilbert Lagrangian to the quadratic Lagrangian.

Nowadays, many new discoveries that might have something to do with gravitation are coming in. 
For example, stellar objects at the spiral arms of the galaxies are rotating at faster speeds than that can be explained by the Einstein's theory. 
To overcome this difficulty, people assume that some extra matter, not visible to us, are giving an extra pull on these stellar objects. 
Extra light deflections, as observed in the intergalactic gravitational lensings, are also ascribed to the existence of these extra matters.
This is the so called $Dark\ Matter$ problem.

An equally well known fact is that the Universe is accelerating in its expansion. 
This is in contrast to the prediction by the Einstein's theory, unless some extra energy is kept pumping into the Universe. 
This is the so called $Dark\ Energy$ problem.

Another longstanding problem facing physicists is the existence of several particle families in the elementary particles realm. 
Corresponding particles in different families differ by only one thing, namely, their inertial masses. 
This is the so called $Family\ Problem$ for the elementary particles.

% \textcolor{red}{**And matter-anti-matter asymmetry??**}

In this letter, we want to show that the quadratic gravitational Lagrangian, when formulated in a form which includes both the metric and the torsion as dynamical variables, will shed some light on the aforementioned problems. 

\section{A story of two metrics}
To meet our goals, we will search for possible solutions, for a spherically symmetric source and also for a homogeneous and isotropic Universe, to the equations of motion of the quadratic Lagrangian.

We will then show, with proper interpretations, that those solutions from a spherically symmetric source can explain the galactic rotations, and the amount of intergalactic lensings.  
We will then also show that the Universe can accommodate acceleration during its expansion if a primordial torsion evolves together with an expanding metric. 
The Family Problem can also be explained as an effect of the various metrics on the energy contents of their generating sources.

We will follow the conventions and notations given in~\cite{Hehl:1979}. The fundamental dynamical variables are the vierbein $e^{\cdot \alpha}_i$ and the connection $\Gamma^{\cdot \alpha \beta}_i$.
Constructed from these variables is the curvature tensor 
\begin{equation}
  F^{\ \ \ \beta}_{ij \alpha} = 2 ( \partial_{[i} \Gamma_{j ] \alpha}^{\ \ \ \beta} + \Gamma_{ [ i | \gamma \ |}^{\ \ \ \beta} \Gamma_{j ] \alpha}^{\ \ \  \gamma} ), 
\end{equation}
and the quadratic gravitation Lagrangian 
\begin{equation}
  \mathcal{L} = \frac{1}{\chi} F_{\alpha \beta \gamma \delta} F^{\alpha \beta \gamma \delta},
\end{equation}
For this Lagrangian, the Schwazschild metric
\begin{equation}
   ds^2 = ( 1 - \frac{2GM}{r})dt^2 - (1 - \frac{2GM}{r})^{-1} dr^2 - r^2 d\Omega^2,
\end{equation}
as well as the Thompson-Pirani-Pavelle metric~\cite{Thomspson:1975, Yeung:1987}, 
\begin{equation}
   ds^2 = ( 1 + \frac{ G'M' }{r})^{-2}dt^2 - (1 + \frac{ G'M' }{r})^{-2} dr^2 - r^2 d\Omega^2,
\end{equation}   
are solutions to the equations of motion, with vanishing torsion.

Unfortunately, this second metric was dismissed by its discoverers soon after its discovery because it failed to reproduce the classical tests that are so successfully predicted by the Einstein-Hilbert Lagrangian.

Here we want to show that a suitable combination of the above two metrics will reproduce nicely both the galactic rotation curves and the amount of intergalactic lensings , if nature follows our following postulate.

We postulate that if nature is going to make use of the quadratic gravitational Lagrangian, matter will then be endowed with either one of these two metrics. 
Matter endowed with the first metric will be called the regular matter, and matter endowed with the second metric will be called the primed matter. 
Regular matter will then interact only with regular matter gravitationally, and primed matter will then interact only with primed matter gravitationally~\cite{EnergyConv}. 
Other than gravitational interactions, regular matter and primed matter are identically the same in all other interactions.

The respective acceleration produced by these metrics, when the speed of motion is small compared with the speed of light, can be calculated from
\begin{equation}
  \frac{d^2 r}{dt^2} = \frac{1}{2}g^{rr}\frac{\partial g_{00}}{\partial r},
\end{equation}
and are respectively $-\frac{GM}{r^2}$ and $-\frac{G'M'}{r^2}( 1 + \frac{G'M'}{r})^{-1}$.

The G' and M' are introduced to make a parallel comparison between the Newtonian gravitational force and the new gravitational force. 
We will call G' the primed gravitational constant, and M' the primed gravitational mass. 
This new Gravitational force is attractive when G'M' is positive. 

\section{Galactic rotation curves}
Because the Gravitational equation of motion is highly nonlinear, a rigorous solution for a source with both regular matter and primed matter together is hard to find. However, we can expect that, for a test object far away from the source, it will see some kind of combination of these two forces.

Hence for a test object, which itself is a mixture of the regular matter of regular mass m and the primed matter of primed mass m', moving under the influence of a distant source of regular matter of regular mass M and primed matter of primed mass M', the combined gravitational forces will be
\begin{equation}
F = -\frac{GM}{r^2} m - \frac{G'M'}{r^2}( 1 + \frac{G'M'}{r})^{-1}m',
  \label{eq:totalF}
\end{equation}

We must be very careful in interpreting this equation. We put the Newtonian force and the new gravitational force side by side simply because we want to emphasize their relative importance in determining the response of the test object at distance $r$. What this equation really mean is that when the Newtonian force is weak, the space-time metric will be that of Thompson-Pirani-Pavelle, and vice versa.

The regular matter and the primed matter in the test object are held together as a single piece by the electromagnet forces between them, and hence will move as a whole. And if it rotates around the source object , the centrifugal force will equal the gravitation force, and hence
\begin{equation}
  (m+m')\frac{v^2}{r} = \frac{GMm}{r^2} + \frac{G'M'}{r^2}( 1 + \frac{G'M'}{r})^{-1}m', 
\end{equation}
and giving a relationship between the rotation speed $v$ with the distance $r$, as
\begin{equation}
  v^2 = \frac{GM}{r}\frac{m}{m+m'} + \frac{G'M'}{r+G'M'}\frac{m'}{m+m'},
  \label{eq:totalV2}  
\end{equation}
Again, we are going to interpret this equation in the way we interpret Eq.[\ref{eq:totalF}].

Note that the speed of light is set equal to 1 here. And also note that both GM and G'M' are of the dimension of length . 

Eq.[\ref{eq:totalV2}] has a salient feature that the Newtonian force hasn't. 
In the scenario where G'M' dominates over $r$, $v^{2}$ has a constant term ($\frac{m'}{m +m'}$) in addition to the Newtonian 1/$r$ term. 
We suspect that this new constant term is what makes the galactic rotation curves flattening when the Newtonian force weakens at the outskirts of the galaxies.

Let us now focus ourselves on a spiral galaxy. 
We will make some assumptions on the mass distribution and regular matter and primed matter composition of the galaxy. 
We will assume that the galaxy consists of a compact disk (a bulge and spiral arms), and an extensive halo. 
The compact disk, which is mostly made up of stars, takes up a large fraction of the total mass of the galaxy. 
While a small fraction belongs to the halo. 
The halo is considered to be diffuse, and uniformly distributed with regular matter with mass density $\rho$ and primed matter with mass density $\rho'$.

The stars in the galaxy are assumed to be devoid of primed matter. 
This assumption can be understood in the following way. 
The primed matter always respond to the primed gravitational pull with a high rotation speed even when they are bound with some regular matter and hence are far harder for them to condense gravitationally into a star.

At the galactic level (a size of tens of kilo parsecs), the halo looks diffuse and uniform. 
But at the level of the size of a few parsecs, the halo could be lumpy. 
These lumps are called dust pockets. Stars, though devoid of primed matter, are bound to their respective neighboring dust pockets, through their Newtonian interactions with the regular matters in the dust pockets. 
The stars and the dust pockets will move, as a whole in the galactic space under the gravitational influence of the matters in the disk and in the halo.

Note that the stars do rotate about the center of their neighboring dust pockets. But such local rotation needs a much smaller velocity to substantiate than that of the galactic rotation because of the small size of the dust pockets.

When the stellar object rotates inside the halo, heuristically we believe that, it will see only the matters that lie inside $r$. 
For the primed matter, the stellar object will see a mass of $\frac{4\pi}{3}\rho'r^3$. 
Note that the contribution to $v^2$ from $\rho$ is negligible because the halo contains only a small fraction of all the regular masses.

So, at a distance of $r$ from the center of the galaxy, the rotation speed of the stellar object (whose mass is small when compared with mass of the accompanying dust pocket) will heuristically look like 
\begin{eqnarray}
v^2 &=& v^2_d(\frac{m}{m+m'}) + \frac{ G'\frac{4\pi}{3}\rho'r^3 }{ r + G'\frac{4\pi}{3}\rho'r^3 }(\frac{m'}{m+m'}) \nonumber\\
    &=& 4\pi G \Sigma_0 h y^2 [I_0(y)K_0(y) - I_1(y)K_1(y) ] \\ 
    && \times (\frac{m}{m+m'}) + \frac{ G^{\star} r^3}{r + G^{\star}r^3 }(\frac{m'}{m+m'})
    \label{eq:totalV2NEW}
\end{eqnarray}
%, where we have anticipated that $m \gg m'$ in the halo. 
The $v_d^2$ is the speed coming from the Newtonian force of the disk~\cite{Toomre:1963} with $y= \frac{r}{2h}$, and the remaining part from the primed matter of the halo. $G^{\star}$ will stand for $G'(\frac{4\pi}{3}\rho')$ here and in our subsequent discussions. 

\begin{figure*}[htbp]
  \begin{center}
    \subfigure[]{
      \label{fig:RC_MilkyWay}
      \includegraphics[width=0.35\textwidth]{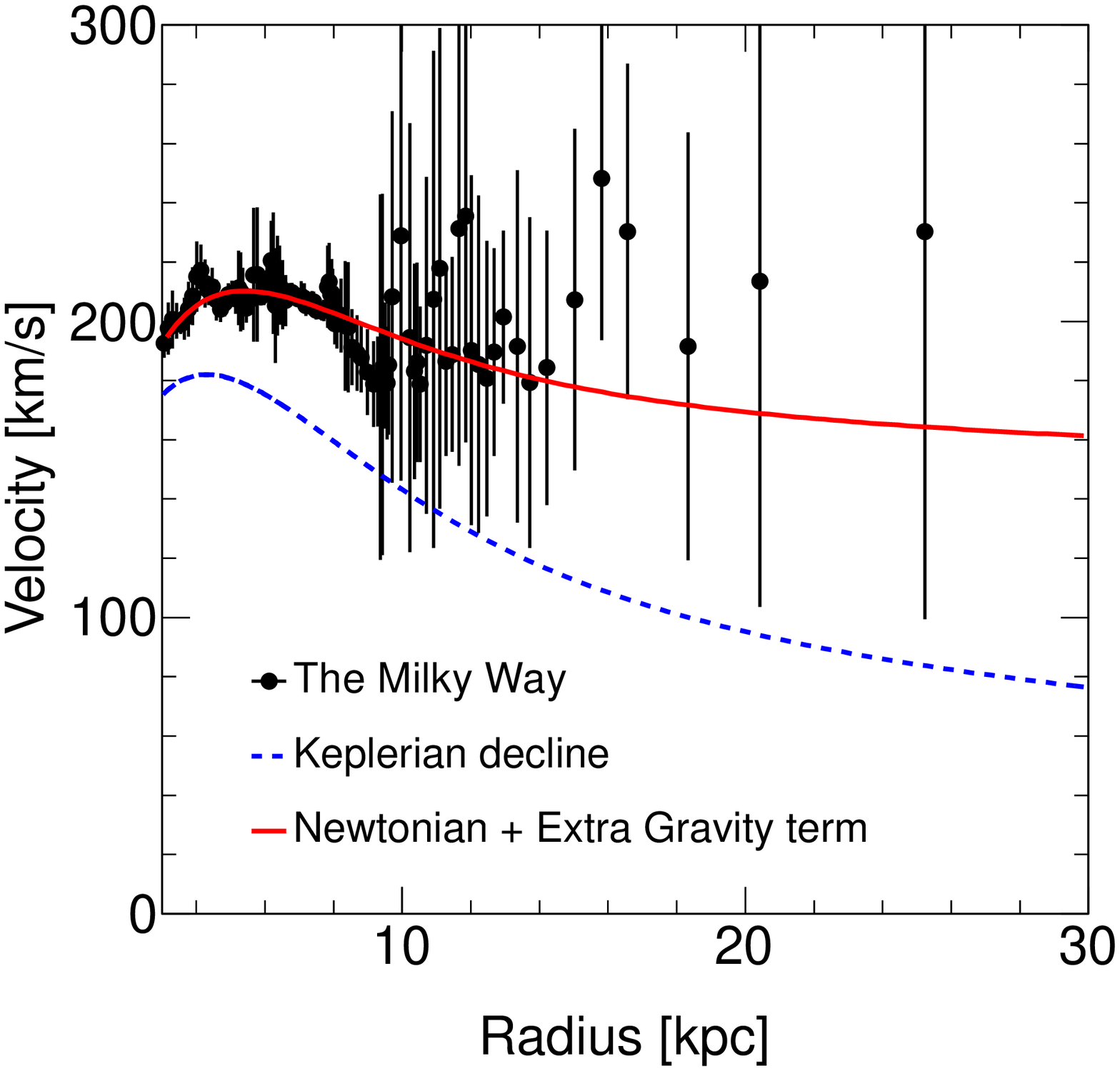}
    }
    \subfigure[]{
      \label{fig:RC_3198}
      \includegraphics[width=0.35\textwidth]{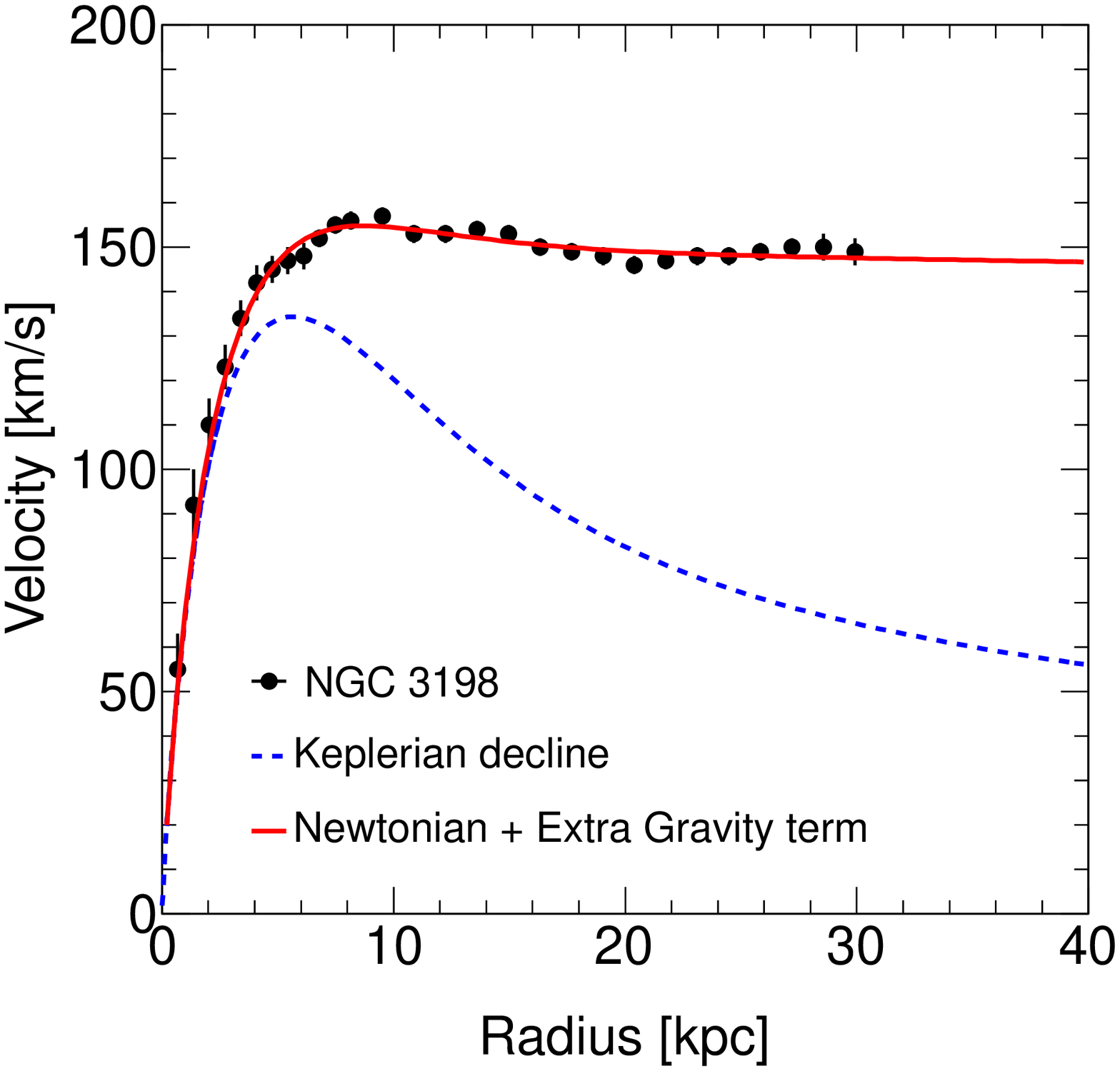}
    }
    \subfigure[]{
      \label{fig:RC_2403}
      \includegraphics[width=0.35\textwidth]{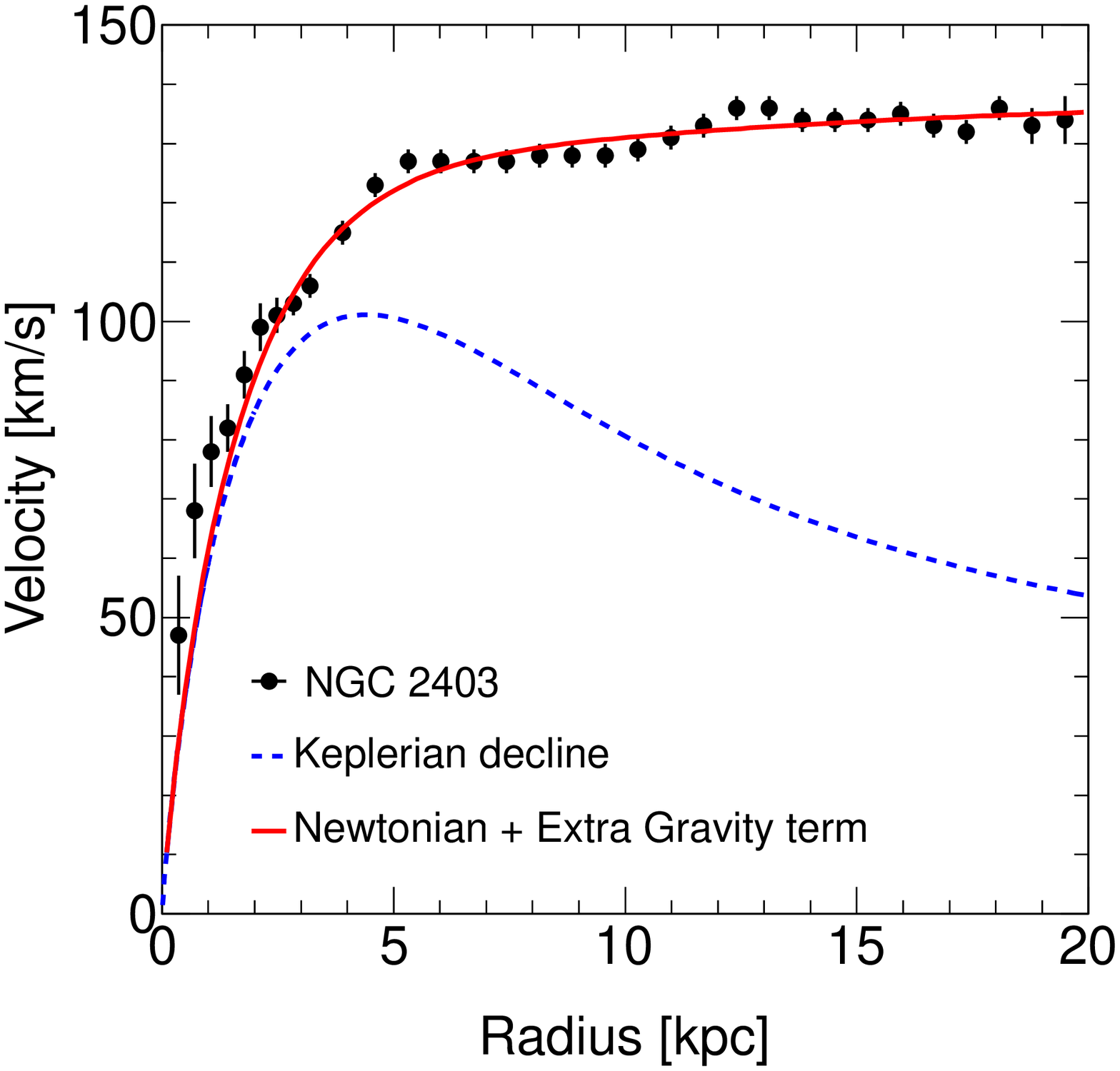}
    }
    \subfigure[]{
      \label{fig:RC_6503}
      \includegraphics[width=0.35\textwidth]{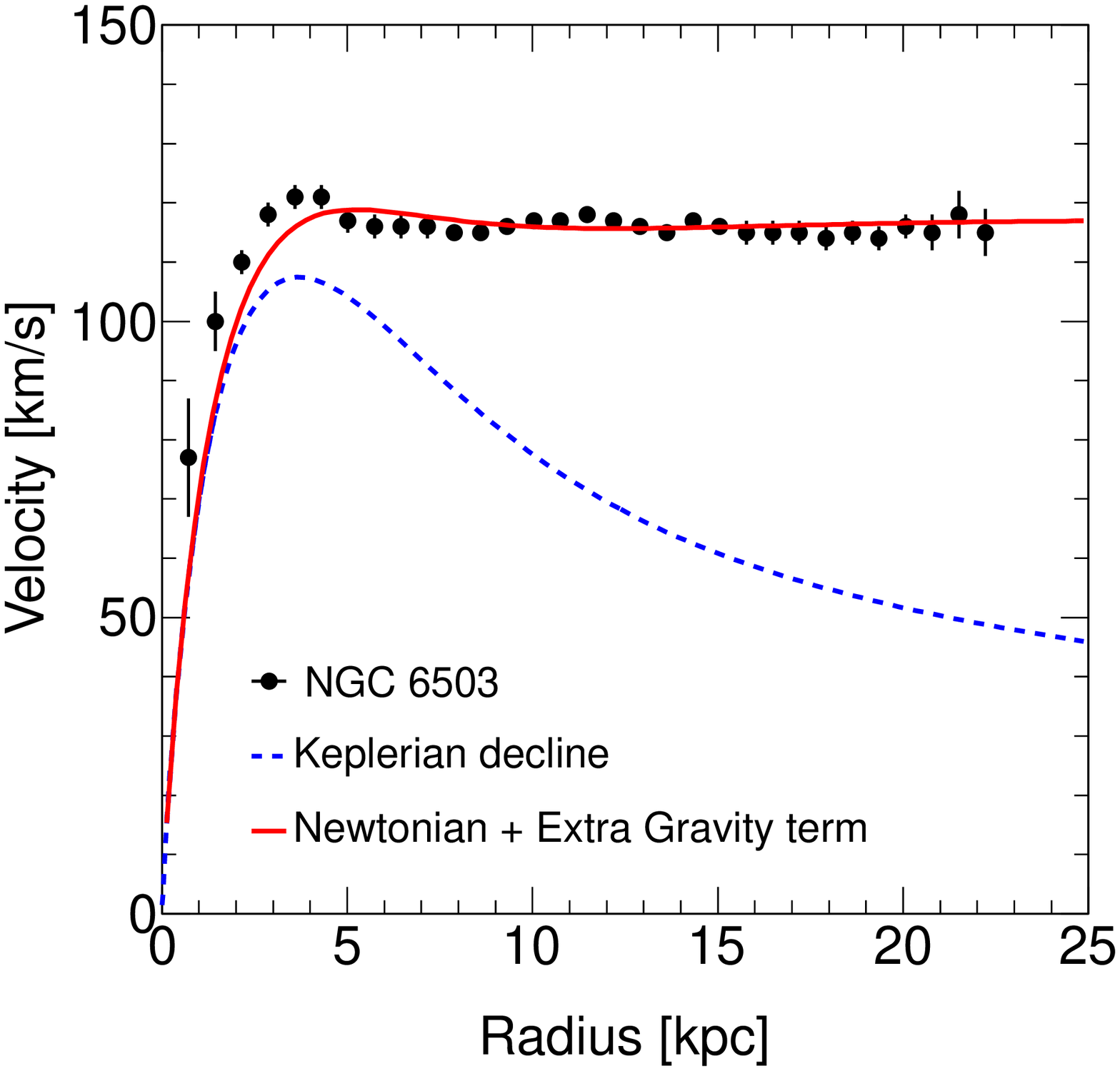}
    }
    \end{center}
  \caption{  The predicted relationship between the galactic rotation speed $v$ and the distance $r$ from a combined influence of the Newtonian force and the new gravitational force: \subref{fig:RC_MilkyWay}The Milky Way, \subref{fig:RC_3198}NGC 3198, \subref{fig:RC_2403}NGC 2403 and \subref{fig:RC_6503}NGC 6503. 
  \label{fig:RC}}
\end{figure*}

The primed gravitational constant $G'$ could be large. And hence $G^{\star}r^3$, could contribute significantly to $v^2$. 
In fact when we try to fit the observed rotation curves of the Milky Way and the galaxies NGC 3198, NGC 2403, and NGC 6503 with that given in Eq.[\ref{eq:totalV2NEW}], we find that the theoretic predictions agree with that observed.
%, a universal value of $G^{\star}$ and a universal value of $\frac{m'}{m+m'}$ give very good fittings of their rotation curves for $r$ ranging from $r$ = 3 kpc to $r$ = 30 kpc. 
Figure~\ref{fig:RC} and Table~\ref{table:RC} give the fitting curves and fitting parameters~\cite{GalaxyData_MW, GalaxyData_1,GalaxyData_2,GalaxyData_3,GalaxyData_4}. 
%\textcolor{red}{**Reference for the data points.**} 

\begin{table*}[htbp]
\large{
   \begin{center}
   \begin{tabular}{c|c|c|c|c}
    \hline
    \hline
                                        &    The Milky Way    &     NGC 3198       &     NGC 2403       &     NGC 6503        \\
    \hline
     $G\Sigma_0$ [$km^2s^{-2}kpc^{-1}$] &  $6.8\times10^{3}$  & $2.8\times10^{3}$  & $2.1\times10^{3}$  &  $2.8\times10^{3}$   \\
      h  [kpc]                          &        2.0          &   2.63             &      2.05          &      1.72            \\
     $G^{\star}$  [kpc$^{-2}$]          &  $5.0\times10^{-2}$ & $9.2\times10^{-3}$ & $1.4\times10^{-2}$ &  $1.3\times10^{-2}$  \\
     $\frac{m'}{m}$                     &  $2.3\times10^{-9}$ & $2.2\times10^{-9}$ & $2.0\times10^{-9}$ &  $1.4\times10^{-9}$  \\
    \hline
     \hline
   \end{tabular}
    \end{center}
   \caption{The fitting parameters for different galaxies.}
\label{table:RC}
}
\end{table*}

It is amazing to find that a universal value of $G^{\star} \approx 10^{-2}\ kpc^{-2}$ and a universal ratio of $\frac{m'}{m} \approx \ 2 \times10^{-9}$ fit very well with the observed results when $r$ ranges from 3 kpc to 30 kpc. Note also that the value of $G\Sigma_0$ and $h$ are more or less the same as those observed.

\section{Intergalactic lensing}
Next, let us turn to see what the primed matter do in explaining the large light deflections that are observed in intergalactic gravitational lensings. 
We shall take the galaxy cluster Abell 1689 as our illustration. 
We shall regard Abell 1689 as a cluster consisting of galaxies which are carrying their own individual halos with them. 
And this collection of galactic halos forms the halo of the cluster. 
Since the galactic halos are always regarded as having a size of the order of 30 kpc, the size of the cluster will be very close to its halo size which is taken to be 300 kpc.

The Azimuthal angle swept by the light, when it travels from point $R$ to the point of closest approach $r_0$, under the influence of gravity described by the metric
\begin{equation}
  ds^2 = B(r)dt^2 - A(r)dr^2 -r^2d\Omega^2,
  \label{eq:WeibEq}  
\end{equation}
is given by~\cite{Weinberg}
\begin{equation}
  \Delta\varphi \equiv \varphi(r_0) - \varphi(R) = \int^{R}_{r_0} A^{\frac{1}{2}}(r)[(\frac{r}{r_0})^2 \frac{B(r_0)}{B(r)} -1 ]^{-\frac{1}{2}} \frac{dr}{r},
  \label{eq:LightDef}  
\end{equation}

In the case of a Thompson-Pirani-Pavelle metric for a point source of primed mass M', the angle swept is
\begin{equation}
  \Delta\varphi = \int^{R}_{r_0} \frac{(r_0 + G'M')dr}{(r+G'M')[(r+G'M')^2 - (r_0+G'M')^2]^{\frac{1}{2}}},
  \label{eq:LightDef_TPP}  
\end{equation}
which can be readily integrated to give
\begin{equation}
  \Delta\varphi = \sec^{-1} \frac{1+\beta R}{1+\beta r_0},
\end{equation}
with $\beta$ = $\frac{1}{G'M'}$.

We will immediately notice that the angle change will be $\pi/2$ when $R$ goes to infinity. 
That means no deflection for the light by a point source of primed matter when it comes from infinity and then goes back to infinity. 
Things will be different when we deal with a distributed source of primed matter as we are going to show in the following.

Let $R$ be the radius of the halo of Abell 1689 and $r_0$ be the closest approach from the cluster center. 
The light will see a point source of primed matter of constant mass M' when it is moving beyond the halo. 
And it will see a point source of diminishing mass (i.e. increasing $\beta$) when it enters the halo because it sees only the mass that lies inside $r$.

We claim that the deflection by a point source of mass M' is smaller than that by a uniformly distributed halo which has a total mass of M', if the light penetrates into the halo during some time in its journey. The above claim is obvious by noting that 
\begin{equation}
  \frac{d(\Delta\varphi)}{d\beta} > 0,
\end{equation}

So if the light penetrates into the cluster halo, the total Azimuthal angle change will be greater than $\pi/2$, and we will see light deflectng towards the center of the cluster.

The actual angle swept, when light comes from the infinity, enters the halo at $R$ and reaches the point of closest approach at $r_0$ is given by 
\begin{equation}
  \frac{\pi}{2} + ( \Delta\varphi|_{halo} - \Delta\varphi|_{point\ source} ),
  \label{eq:LightDef_Final}    
\end{equation}

The second term in the bracket in Eq.[\ref{eq:LightDef_Final}] comes from the fact that we have to take out a $\pi/2$ in calculating the angle of deflection.

The quantity inside the bracket of Eq.[\ref{eq:LightDef_Final}] can be regarded as a change $\delta(\Delta\varphi)$ in $\Delta\varphi$ due to a change $\delta M'$ in M'. 
And we get  
\begin{eqnarray}
 &&\Delta\varphi|_{halo} - \Delta\varphi|_{point\ source} \nonumber \\ 
  && = \delta(\Delta\varphi) \nonumber\\
  && = \frac{\partial (\Delta\varphi)}{\partial M'}\delta M'   \\
  && = \frac{\partial}{\partial M'} [\sec^{-1} \frac{1+\beta R}{1+\beta r_0}] \delta M' \nonumber \\ 
  && = \frac{3}{\sqrt{2}} (R - r_0)^{ \frac{3}{2}} G^{\star -\frac{1}{2}} R^{-\frac{5}{2}}  \nonumber 
    \label{eq:DeltaPhi}
\end{eqnarray}

A plug in the data of $R$ = 300 kpc, $(R - r_0)$ = 30 kpc and $G^{\star}$ = $10^{-2}$ $kpc^{-2}$, will give us an angle of deflection, which is 2$\delta (\Delta \varphi)$, a value of $4\times 10^{-3}$, which is many times of that expected from General Relativity.
 
Note that the value of $G^{\star}$ that we used in calculating the Abell 1689 light deflection comes from the curve fittings of galactic rotations. 
Again there seems to be a universal value for $G^{\star}$, as it should be.

We should also note that the sun and the planets in the solar system aren't contaminated with the primed matter as we have explained at the beginning of the letter. 
The fundamental tests on General Relativity will hence remain intact . 

%\begin{figure}[htbp]
%  \begin{center}
%      \includegraphics[width=0.47\textwidth]{RotationCurve_Solar.eps}
%      \label{fig:RC_Solar}
%  \end{center}
%\end{figure}

\section{Particle families}
As we have put in our postulate, the regular matter and the primed matter have the same electroweak and strong interactions, and they differ only in their endowed metrics. 
And as the gravitational energy contents of matter depends on the form of the metrics that go along with them~\cite{Arnowitt:1961}, the regular matter and the primed matter will behave the same in their electroweak and strong interactions except that they may carry different energy contents .
 
This reminds us of the existence of different particle families in our laboratories.
We here speculate that the microscopic regular matter and the microscopic primed matter are the same matter belonging to different particle families. Hence the existence of the various particle families is precisely a reflection of the existence of these various microscopic metrics arising from the solutions of the same quadratic gravitational Lagrangian.
%\textcolor{red}{****How about quark-anti-quark asymmetry??****}

\section{A primordial torsion and the accelerating expansion of
the Universe}
There is another nice feature of the quadratic Lagrangian, when we go to the Universe as a whole. 
The quadratic gravitation Lagrangian was shown to admit a cosmological solution of the form~\cite{Yeung:1985, HsuYeung:1985}
\begin{equation}
 ds^2 = dt^2 - \rho_0^2 e^{2\xi t} ( d\rho^2 + \rho^2 d\Omega^2), 
\end{equation}
with primordial torsion compoments, 
\begin{equation}
  F_{011} = F_{022} = F_{033} = \xi. 
\end{equation}

The $\rho_0$ and $\xi$, with $ \xi > 0$ are integration constants arising from the integration of the equation of motion. 

We can interpret this solution as representing an expanding and accelerating Universe when the influence of gravity dominates over the influence of matters.

The role played by the primordial torsion is crucial here: the stretching on the Universe by the metric is compensated by the twisting by the torsion. 
And from the metrical point of view, we look like living in a Universe with a cosmological constant $\xi$.  

And interesting enough, our torsion selects the spatially flat metric ($\kappa$ = 0) as the only accompanying metric~\cite{Yeung:1985, HsuYeung:1985}.
The spatially flat geometry of the Universe is confirmed by WMAP.

Nature seems enjoying the full use of the quadratic gravitational Lagrangian.

%%%%%%%%%%%%%%%%%%%%%%%%%%%%%%%%%%%%%%%%%%%%%%%%%%%%%%%%%%%%%%%%%%%%%%%%%%%%%%%
% Acknowledgments
%%%%%%%%%%%%%%%%%%%%%%%%%%%%%%%%%%%%%%%%%%%%%%%%%%%%%%%%%%%%%%%%%%%%%%%%%%%%%%
\section{Acknowledgments}
Yi Yang would like to thank for the full support from Professor Harold Evans (Indiana University).

%%%%%%%%%%%%%%%%%%%%%%%%%%%%%%%%%%%%%%%%%%%%%%%%%%%%%%%%%%%%%%%%%%%%%%%%%%%%%%%
% Bibliography
%%%%%%%%%%%%%%%%%%%%%%%%%%%%%%%%%%%%%%%%%%%%%%%%%%%%%%%%%%%%%%%%%%%%%%%%%%%%%%
%\clearpage
%\bibliographystyle{atlasstylem}
%\bibliography{a1MuMu}

%\bibliographystyle{abbrv}
%\bibliographystyle{unsrt}
%\bibliography{AccUniverse}

\end{document}